# Intersubband Quantum Disc-in-Nanowire Photodetectors with Normal-incidence Response in the Long-wavelength Infrared

*Mohammad Karimi,[1,2] Magnus Heurlin,[1,3] Steven Limpert,[1] Vishal Jain,[1,2] Xulu Zeng,[1] Irene Geijselaers,[1] Ali Nowzari,[1] Ying Fu,[4] Lars Samuelson,[1] Heiner Linke,[1] Magnus T. Borgström[1] and Håkan Pettersson[1,2*]*

[1]Solid State Physics and NanoLund, Lund University, Box 118, SE-221 00 Lund, Sweden

[2]Department of Mathematics, Physics and Electrical Engineering, Halmstad University, Box 823, SE-301 18 Halmstad, Sweden

[3]Sol Voltaics AB, Lund, Sweden

[4]Department of Applied Physics, Royal Institute of Technology (KTH), Science for Life Laboratory, SE-171 21 Solna, Sweden

**Abstract**

Semiconductor nanowires offer great potential for realizing broadband photodetectors that are compatible with silicon technology. However, the spectral range of such detectors has so far been limited to selected regions in the ultraviolet, visible and near-infrared. Here, we report on broadband nanowire heterostructure array photodetectors exhibiting a photoresponse from the visible to long-wavelength infrared. In particular, the infrared response from 3-20 μm is enabled by normal incidence excitation of intersubband transitions in low-bandgap InAsP quantum discs synthesized axially within InP nanowires. The optical characteristics are explained by the excitation of the longitudinal component of optical modes in the photonic crystal formed by the nanostructured portion of the detectors, combined with a non-symmetric potential profile of the discs resulting from synthesis. Our results provide a generalizable insight into how broadband nanowire photodetectors may be designed, and how engineered nanowire heterostructures open up new fascinating opportunities for optoelectronics.

**Introduction**

Nanowires (NWs) are versatile optoelectronic device building blocks[1, 2, 3, 4] since their optical and electronic properties can be tailored in unique ways.[5, 6] Furthermore, strain relaxation allows them to host defect-free lattice-mismatched heterostructures[7] and to grow III-V semiconductors directly on silicon[8], which makes them promising candidates for next-generation III-V-on-silicon photodetectors in optical communications, surveillance and thermal imaging. However, the spectral range of NW-based photodetectors has so far been limited to selected regions of the ultraviolet, visible and near-infrared.[9] Recently, a few reports were published on a photoresponse extending into the mid-wavelength infrared (MWIR, 3-8 μm).[10, 11] Presently, the longest operation wavelength reported for a NW-based photodetector is approximately 5.7 μm, which was demonstrated in a low-bandgap InAsSb $p^+$-i-$n^+$ NW array[11] that exhibited a photoresponse peak at 4 μm with a roll off towards both shorter and longer wavelengths.

Planar photodetectors for the long-wavelength infrared region (LWIR, 8-15 μm) are based on interband (band-to-band) transitions in either $Hg_xCd_{1-x}Te$ (MCT) alloys[12, 13] and InAs/GaSb type-II superlattices (T2SLs)[14], or on intersubband transitions between subbands in the conduction or valence band of quantum well (QW) infrared photodetectors (QWIPs).[15] Each of these LWIR detectors have their own strengths and weaknesses. For example, MCT detectors are superior to QWIPs in terms of detectivity and operation temperature. However, MCT detectors compare poorly to QWIPs in terms of fabrication uniformity and yield, and they contain toxic elements.

Despite the rapid pace of NW photodetector development, and the mature nature of planar technology for LWIR photodetection, there are so far no reports of either LWIR absorption or intersubband photocurrent generation in NWs. Here, we report on NW photodetectors capable of broadband detection at normal incidence in the MWIR and LWIR range, extending to wavelengths as long as 20 μm. The broadband detection is a result of intersubband photocurrent generation in arrays of oxide-capped, photoresist-embedded $n^+$-i-$n^+$ InP NWs with single- or multiple axial InAsP quantum discs (QDiscs). The photoresponse of our devices extends from the short-wavelength part of the far-infrared region (FIR, 15 μm – 1mm) to the visible, demonstrating that NWs are a viable platform for broadband detection. Furthermore, the NWs exhibit photonic crystal properties, thus enabling new opportunities controlling and manipulating long-wavelength light in detection and energy-harvesting devices.

**Results**

The NW photodetectors consist of arrays of 4 million InP NWs grown by metal-organic vapor phase epitaxy (MOVPE) on InP substrates with single or multiple (20) InAsP low-bandgap quantum discs (QDiscs) grown axially in each NW (Figure 1a-c). Figure S1 in the Supplementary Information shows the photoluminescence (PL) characteristics of a single NW with an embedded QDisc. At low excitation intensity, a ground state PL peak originating from the QDisc is observed at about 0.90-0.95 eV. The QDisc PL signal increases and blueshifts with increasing excitation density, which we attribute to complex recombination transitions between different electron and hole states. Large array photodetectors were subsequently fabricated by coating the NW arrays with $SiO_2$ and embedding them in a dielectric matrix and adding contacts (Figure 2a-f). In-depth current-voltage (I-V) characterization and spectrally-resolved photocurrent (PC) measurements, using Fourier transform infrared (FTIR) spectroscopy, were pursued on the devices at different temperatures (see Methods section for experimental details).

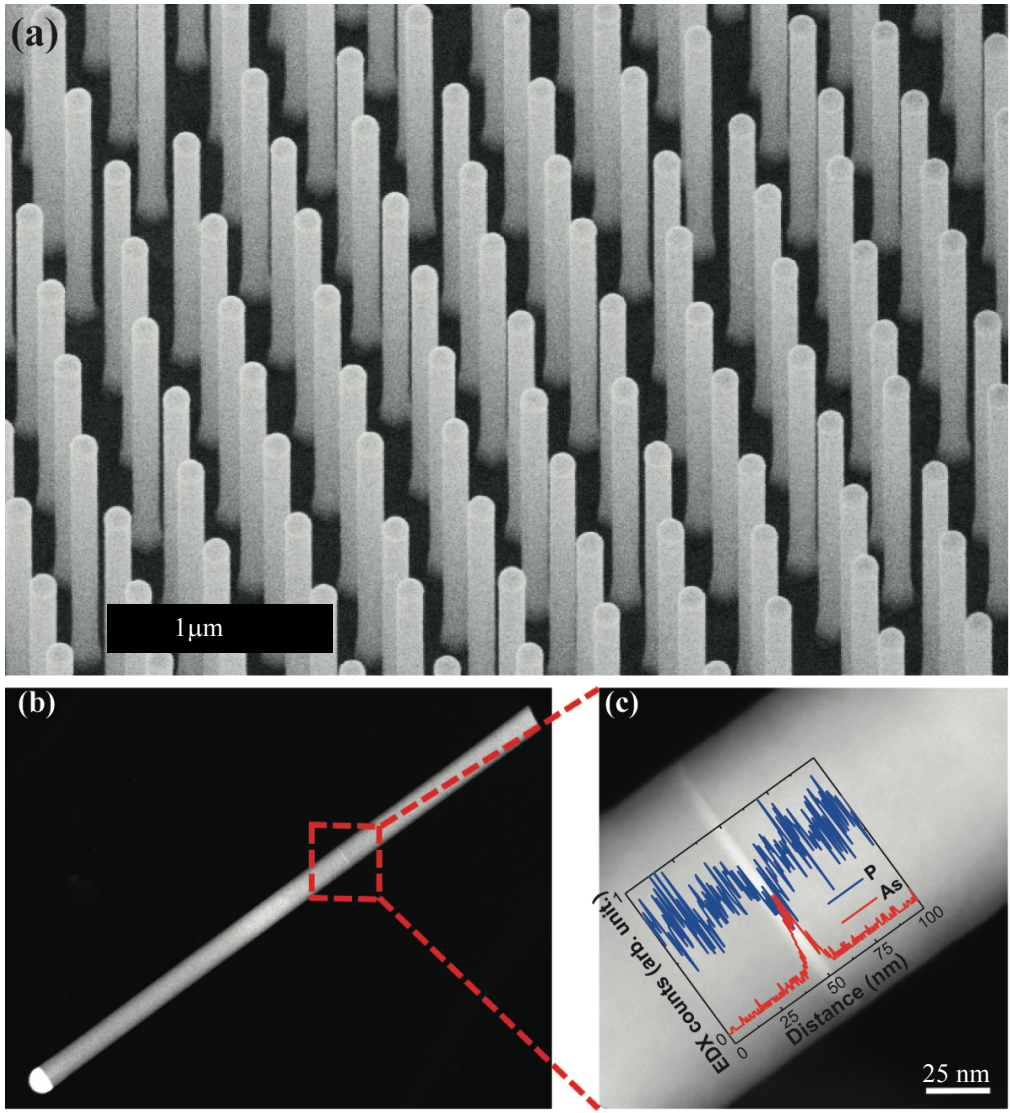

**Figure 1.** (a) SEM image of an as-grown InP NW array with a single embedded InAsP QDisc in each NW. (b) Low-resolution STEM image of a NW with an embedded disc. (c) A higher resolution image showing the QDisc, overlaid with an EDX linescan. The EDX profile show that the discs are non-abrupt with an average thickness of 7-8 nm. Atomic concentrations of 55% As and 45% P were extracted from point scan measurements in the middle of the QDisc.

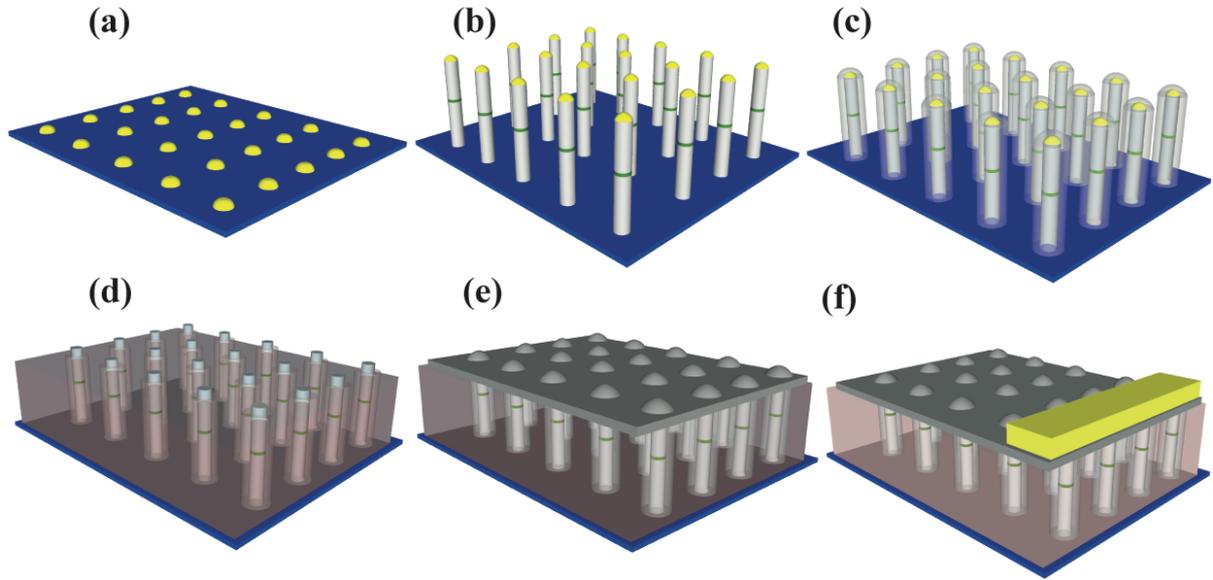

**Figure 2.** The sequence of steps used to grow the NW arrays and create the photodetector devices. (a) Patterning of catalytic gold particles on InP substrates by NIL. (b) MOVPE growth of NWs with embedded QDiscs. (c) Deposition of a $SiO_x/Al_2O_3$ bi-oxide layer using ALD. (d) Spin-coating of a thick photoresist spacing layer between the NWs, followed by sequential dry back-etching to reveal the tips of the NWs. Etching of $SiO_x/Al_2O_3$ at the tips (top contact points) of the NWs and the gold particles. (e) Connecting the tips of the NWs by sputtering ITO, followed by (f) Ti and Au evaporation to define a bond pad.

InP NWs typically show a weak n-type behavior due to residual dopants, which increases the dark current of n-i-n photodetectors. To overcome this, we have used Zn co-doping during growth of the nominal i-segment. This compensates for residual n-doping and is thus effective in suppressing the dark current (Figure 3a). This results in a dark current density that is comparable to conventional optimized quantum dot and superlattice infrared photodetectors.[16, 17, 18] In brief, the low dark current levels – falling by about 5 orders of magnitude from $4\times10^{-4}$ A at 300 K to $5\times10^{-9}$ A at 5 K for a bias of 0.5 V – and the symmetric I-V characteristics indicate that we have been successful in fabricating high-quality photoconductive detector elements free of radial self-gating[19] and Zn overcompensation[19].

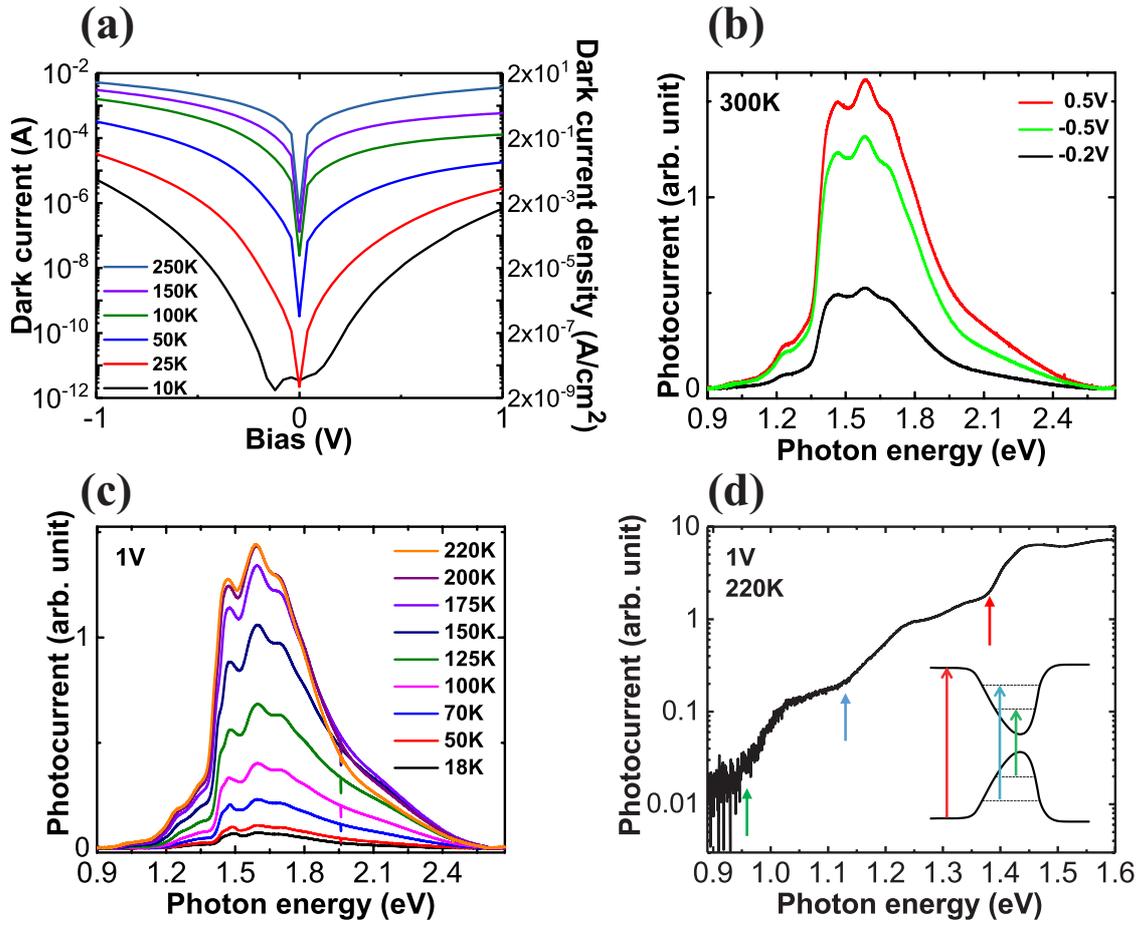

**Figure 3.** Current-voltage characteristics and spectrally resolved interband PC of a QDisc-in-NW detector. (a) I-V characteristics of an array detector measured at different temperatures. (b) PC at 300 K for different applied biases. (c) PC at 1 V bias versus temperature and (d) PC at 220 K, taken from (c), in semi-log scale. Inset shows the relevant QDisc interband transitions schematically.

The spectrally-resolved PC response shows that the fully processed device exhibits a broad spectral response across visible and near infrared wavelengths from 0.9 to 2.4 eV (0.5 to 1.4 μm) at room temperature (Figures 3b-d). We attribute the PC response in the 1.3 to 2.4 eV range to interband transitions in the InP and in the 0.9 to 1.3 eV range to interband transitions between confined energy levels in the InAsP QDisc. The semi-log plot of the PC in Figure 3d clearly indicates two interband transitions with onset of about 0.95-1.0 eV and 1.1-1.15 eV, respectively. The lowest energy (ground state) interband transition in the QDiscs is in good agreement with the single NW PL in Figure S1. The inset in Figure 3d shows the assignment of the corresponding underlying optical transitions. The proposed smooth and asymmetric potential of the discs is inferred from the EDX data in Fig. 1c. Strong peaks in the energy range between 1.45 to 1.65 eV are interpreted as interband transitions in the zincblende (ZB) and wurtzite (WZ) segments of the polytype InP NWs, consistent with previously reported photocurrent response in InP NWs.[20] Temperature-dependent PC spectra (Figure 3c) show that the peaks develop significantly with increasing temperature, which can be explained by enhanced carrier collection at elevated temperatures due to reduced trapping by the staggered conduction band landscape induced by the polytype ZB/WZ InP crystal structure[21] and the embedded QDiscs – a characteristic that has been observed in other NW-based photodetectors[19, 22] and in quantum ring infrared photodetectors.[23] In addition to signal enhancement with increasing temperature, the PC spectra are slightly red-shifted because of the decreased band gap of the InP NWs and InAsP QDiscs.

At low temperature, spectrally-resolved PC spectra show that the devices exhibit a PC response that extends out to approximately 20 μm and is tunable via the QDisc design. The PC spectra of the single-QDisc device (Figure 4a) shows a large PC signal centered around 0.1 eV at normal incidence, having an apparent double-peaked structure with peaks at approximately 0.08 and 0.12 eV. Additionally, the single-QDisc exhibits two smaller photocurrent peaks at 0.20 eV and 0.40 eV. In comparison, the multiple-QDisc device (Figure 2 in SI) shows a broader photocurrent signal extending from approximately 0.1 eV to 0.35 eV. These photocurrent signals in the MWIR and LWIR regions were recorded after switching from interband (CaF$_2$ beam-splitter and quartz lamp) to intersubband (KBr beam-splitter and globar filament source) spectrometer settings. We observe that the intensity of the intersubband PC signal is largest at normal incidence, and is reduced as the angle of incidence increases (Figure 4b). This behavior is in sharp contrast to conventional planar QW detectors that exhibit very small PC signals at normal incidence due to fundamental selection rules for intersubband transitions as discussed below. In addition, the PC signals in our detectors reduce at different rates for different wavelengths.

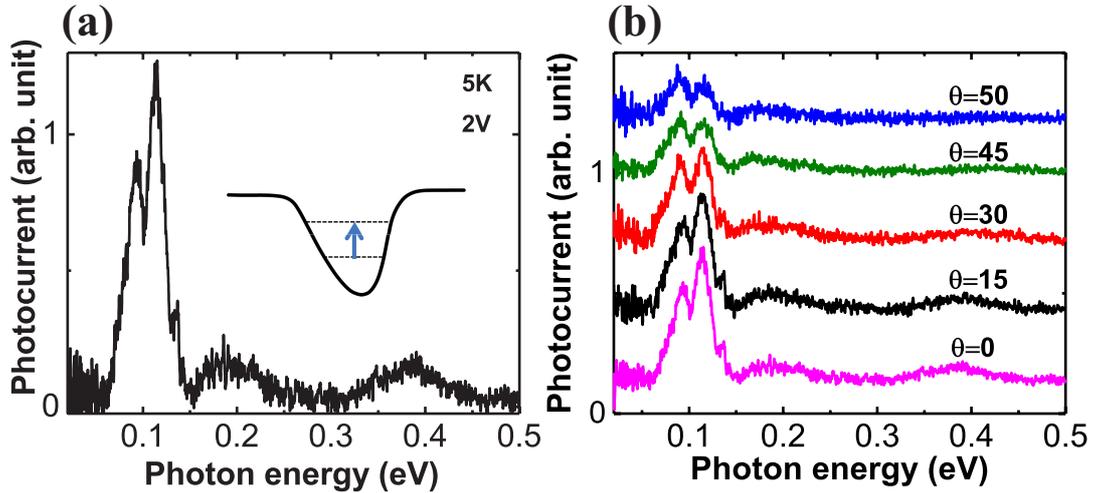

**Figure 4**. (a) Spectrally-resolved intersubband PC at normal incidence excitation. The inset shows schematically the electronic structure of the discs with indicated energy levels and the main optical transition behind the experimentally observed signal at about 0.1 eV. (b) PC spectra taken at various angles of incidence relative to the normal of the sample.

**Discussion**

In the following sections, we first discuss the electronic origin of the intersubband transitions that is responsible for our devices' long-wavelength PC response. Then, we discuss how variations in growth parameters lead to differences between the single-QDisc and the multiple-QDisc intersubband photoresponse. Finally, we discuss the optical origin of our observation of intersubband transistions under normal incidence illumination – a condition in which intersubband transitions cannot be excited in planar QWIPs. In this latter section, we detail the optical properties of our NW photonic crystal photodetectors that make them fundamentally different from planar intersubband detectors and from other NW-based photodetectors.

The electronic origin of the intersubband transitions in our single-QDisc device can qualitatively be understood from calculations of the band structure of the InAsP QDiscs using an 8-band $k \cdot p$ model,[24] which shows that there are two confined energy levels in the conduction band of a single QDisc. Since the QDiscs (NWs) have a fairly large diameter of 130 nm and a nominal thickness of 7-8 nm, we apply a conventional planar QW model that neglects any radial confinement. Since the detailed electronic parameters e.g. effective masses of electrons and holes and band alignment for WZ InAs are unknown, we assume in the calculations that the discs have a ZB structure. Our calculations show the existence of two states inside the conduction band of the $InAs_{0.55}P_{0.45}$ QDiscs, a ground state at 0.22 eV and an excited state at 0.05 eV below the conduction band of the InP. As evident from the EDX data in Figure 1c, the As composition varies significantly across the thickness of the QDiscs which leads to a smoothened potential profile (see inset of Figure 3d). In effect, this pushes up the ground state and lowers the excited state of the QDisc compared to a hard-wall QW potential. The PL data in Figure S1 in the Supplementary Information also supports non-abrupt QDiscs with significant broadening observed at increasing laser excitation. Based on these arguments, the double-peak structure centered around 0.1 eV can be attributed to a conduction band intersubband transition between the QDisc´s ground state and excited state. Taking into account an expected smaller intersubband splitting in the valence band of less than 50 meV, this agrees with the approximate interband splitting of 150 meV deduced from Figure 3d. The discrepancy between our theoretical and the experimental results can be explained by uncertainties in (1) the strain distribution of the large diameter discs, (2) the thickness of the large QDisc ensemble, (3) the composition distribution along the growth direction of the discs (as the invoked deplete mode growth is governed by supply of In stored in the gold catalyst particle)[25] and (4) the electronic parameters of WZ InAsP (see discussion below). The origin of the dip at 0.1 eV (12.4 μm) between the peaks at 0.08 and 0.12 eV is attributed to strong optical phonon absorption centered in the $SiO_2$[26] shell around the NWs, which decreases the amount of light that reaches the QDisc. The PC signals at 0.20 eV and 0.40 eV can be attributed to photoionization of electrons (possibly holes) from the disc´s ground state into the continuum conduction (valence) band of the InP. A small contribution from holes to the measured PC signal in our nominal $n^+$-i-$n^+$ detectors is in fact possible due to the fact that we employ a Fourier transform spectrometer with simultaneous presence of photons with energies up to about 1.3 eV. The high energy photons can thus excite interband transitions in the discs, effectively loading the discs with both electrons and holes.[27] Between 0.25 eV and 0.40 eV the CVD diamond window in the cryostat exhibits a broad absorption peak which at least partly explains the apparent existence of two separate PC peaks. A general concern in optoelectronics is the reduced transmission of ITO at long wavelengths. In our devices, the transmission of the thin ITO layer is about 30% at 0.10 eV, extracted from transmission experiments on ITO-coated semi-insulating InP substrates.

The differences between the intersubband photocurrent signals in the single-QDisc and the multiple-QDisc device may be attributed to variations in the NW growth conditions that affect the QDisc geometry, composition and band structure. Specifically, the multiple-QDisc device exhibited thinner QDiscs in comparison to the single-QDisc device as the growth time for the discs was deliberately reduced. The thinner discs result in higher confined energy levels and, therefore, a blue-shifted intersubband photoresponse. Additionally, the multiple-QDisc device exhibited a larger variation in QDisc thickness and composition than the single-QDisc device because of its deplete-type of growth involving fast switching of the supply of growth precursors. Variations in the thickness and composition of the QDiscs results in varying energy levels, and so a broadening of the intersubband photoresponse.

Optical modeling of a complete single-QDisc device have been done using 2D finite-difference eigenmode (FDE) and 3D finite-difference time domain (FDTD) device models. Interestingly, these show that our detector is a photonic crystal that hosts optical modes with non-zero longitudinal components that extend out of the low-index dielectric matrix and into the high-index NWs. Our observation of intersubband transitions at normal incidence is surprising, and is sufficiently unusual to require a physical explanation, not least because planar QWIPs exhibit zero photocurrent at normal incidence. This is because electromagnetic waves are transverse waves in free space and in thin film stacks, oscillating perpendicular to their direction of propagation. Therefore, when a planar QW intersubband detector is illuminated at normal incidence there is no electric field component in the direction of the QW confinement potential – a requirement for exciting an intersubband transition.[28] In contrast to the physics of these planar devices, eigenmode solutions to Maxwell's equations on a 2D cross-section of our device show that it supports optical modes having non-zero longitudinal components that extend from the low-index dielectric matrix into the high-index NW. These modes ensure that even at normal incidence, the component of the electric field necessary to excite intersubband transitions within the QDiscs is present within the NW. To obtain detailed information about the optical modes present in our device, we used the Lumerical MODE finite-difference eigenmode (FDE) solver. In MODE, we constructed a 2D cross-section unit cell of the device, including the photoresist, oxide and NW, using complex refractive index data for the Microposit S1800 G2 Series Photoresist, $SiO_2$[29] and InP.[29] We applied periodic boundary conditions at the unit cell sidewalls. Eigenmode solutions showed that the structure hosts two optical modes in the 3 to 20 μm wavelength range. Because of variations in the refractive index of $SiO_2$ across this wavelength range, the exact modal shape varies with wavelength. However, two common factors found for the two modes across this wavelength range are that the light is confined primarily within the oxide shell of the NW, with some leakage into the photoresist, and that the modes exhibit an unexpected non-zero longitudinal (i.e. $E_Z$) component that extends into the NW (Figure 5a-d). The 2D FDE model also shows that the modal shape, effective index and longitudinal component amplitude all change with wavelength. The variation in modal shape and effective index means that different wavelengths will have different coupling efficiencies into their respective modes. The variation in longitudinal component amplitude indicates that the intersubband absorption strength may vary with wavelength. For example, the FDE model shows that the longitudinal component amplitude at 10.3 μm (0.12 eV) is significantly greater than at 15.5 μm (0.08 eV) (Figure 5c and d), in agreement with our experimental observation that the PC signal is significantly stronger at 0.12 eV than at 0.08 eV at normal incidence (Figure 4a). The presence of optical modes in the low-index regions of the structure indicates that our device is a photonic crystal. This strongly differentiates our device from planar QW intersubband detectors, which do not employ optical modes, and from other NW-based photodetectors, which employ fiber-type guided modes in the high-index NW "core". [30]

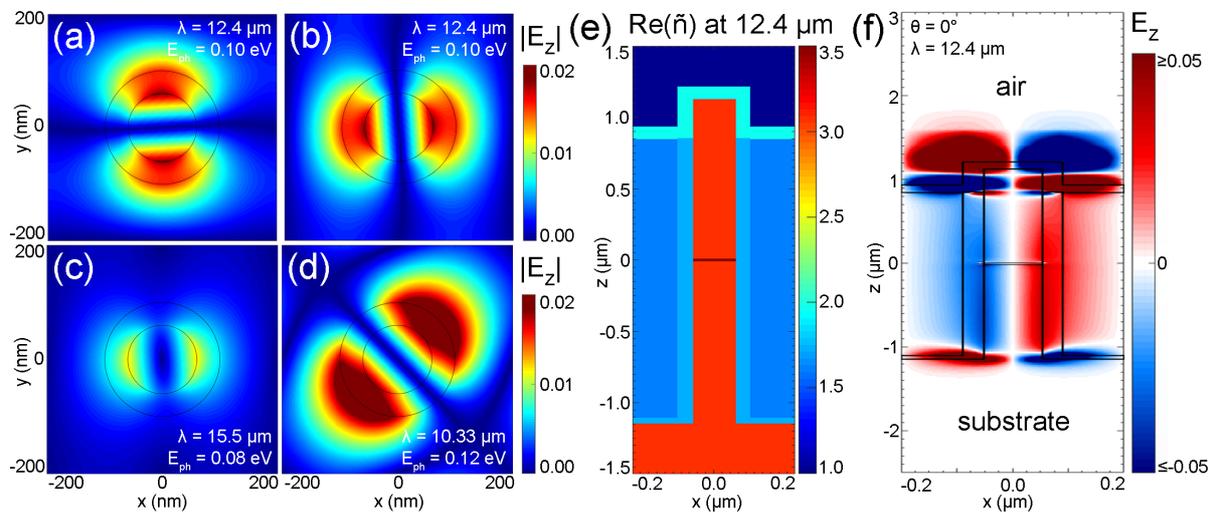

Figure 5. 2D FDE model, 3D FDTD model, and the results of EM modeling. (a)-(b) Amplitudes of the longitudinal component, $|E_Z|$, of the first and second mode at λ = 12.4 μm, respectively. (c)-(d) Amplitudes of the longitudinal component, $|E_Z|$, of the first mode at λ = 15.5 μm and of the second mode at λ = 10.33 μm, respectively. In (a) – (d) the edges of the InP and SiO$_2$ regions are indicated by black lines. (e) 2D cross-section of the 3D unit cell in the x-z plane showing the real part of the complex refractive index at λ = 12.4 μm. Colors indicate the InP substrate and wire (red), the QDisc (dark red), the oxide (light blue), the photoresist (dark blue), the ITO (aqua), and air (dark purple). (f) A slice in the x-z plane of the 3D FDTD model of the longitudinal component, $E_Z$, of the electric field for λ = 12.4 μm at normal incidence (θ = 0).

To evaluate the optical properties of the full 3D device, we used Lumerical FDTD. A 3D unit cell of the device was constructed, including the substrate, the NW with a single QDisc, the $SiO_2$ layer, the photoresist planarization layer, and the ITO top contact. Periodic boundary conditions were applied at the unit cell sidewalls and perfectly matched layers (PMLs) were utilized at the unit cell top and bottom boundaries. The $InAs_{0.55}P_{0.45}$ QDisc was approximated in the model as an InAs QDisc because refractive index data for the $InAs_{0.55}P_{0.45}$ alloy was not available. Complex refractive indices for InP[29], InAs[31], $SiO_2$[29], ITO[32] and the Microposit S1800 G2 Series Photoresist were used (Figure 5e). The 3D FDTD model supports the finding that our device hosts guided optical modes in the oxide shell with non-zero longitudinal components that extend into the NW and disc. A slice of the $E_Z$ component of the electric field for a wavelength of 12.4 μm at normal incidence (Figure 5f) shows that the wave acquires an $E_Z$ component that extends into the NW as it passes through the nanostructured layer of the device, and that this $E_Z$ component vanishes as the wave passes into the substrate.

The presence of a finite $E_Z$ component induced by the nanostructured parts of a photonic crystal at least partly explains the intriguing existence of a normal incidence sensitivity of our NW array detectors. An additional effect which influences the sensitivity to normal incidence radiation is that the compositional variation along the discs (as extracted from the EDX data in Figure 1c) results in a smooth, asymmetric potential that lacks inversion symmetry (as indicated in the inset of Fig 4a). This compositional variation most likely reflects As carry over[33, 34] which leads to a significantly sharper transition from InP to InAsP than that from InAsP to InP. Moreover, our QDiscs are grown using the deplete growth mode (explained in the Methods section) which might induce further compositional changes compared to conventional planar growth. The absence of inversion symmetry of the confining potential washes out the definite parity of the corresponding wavefunctions and leads to non-vanishing oscillator strength in the intersubband transitions at normal incidence.

While the results of the optical modeling provide novel insight into how light is coupled into the discs at normal incidence, it does not account for the strong decrease of intersubband PC on angle of incidence shown in Figure 4b. Instead, we attribute this decrease to a strongly reduced optical interband excitation and loading of the discs with electrons and holes as discussed above. Indeed, angle-dependent PC measurements on QDisc-in-NW photodetectors have clearly indicated a strongly decreasing interband PC signal from the discs with increased angle of incidence. In fact, this decrease follows a similar angle-dependence as the interband signal in InP NW photodetectors/solar cells without QDiscs.[20]

**Conclusion**

We have realized photonic crystal photodetectors comprising 4 million $n^+$-i-$n^+$ InP NWs containing single- or multiple (20) InAsP axial QDiscs which exhibit a photoresponse from the visible to the far-infrared region. We report, for the first time, electronic detection of long-wavelength infrared absorption in NWs and PC originating from intersubband transitions in NW heterostructures. Additionally, we observe PC caused by intersubband transitions under normal incidence illumination, which we attribute to optical modes present in our structure having non-zero longitudinal components that extend out of the low-index dielectric matrix and into the high-index NWs. In addition, an asymmetric disc profile, correlated to an As-carry over effect, relaxes the selection rules for intersubband transitions. Our experimental and modeling results show that NWs are capable of broadband detection, and that integration of photonic crystals and NWs offers new opportunities to control and manipulate long-wavelength light in detection and energy-harvesting devices.

**Methods**

For NW growth, n$^+$-InP (111)B substrates were patterned with Au particles by nanoimprint lithography (NIL), metal evaporation and lift-off. The final pattern comprised 20 nm thick Au discs with a diameter of 180 nm that, under growth conditions, formed Au-In alloy particles 130 nm in diameter. The center-to-center distance (pitch) between two Au-In particles was 400 nm. The growth was carried out in a low-pressure (100 mbar) Aixtron 200/4 MOVPE at 440 °C. The grown NWs had a diameter of 130 nm and a length of about 2.3 μm. In order to compensate for the typical residual n- doping ranging from $10^{15}$ to $10^{16}$ cm$^{-3}$ that is obtained in NWs grown by MBE[35] and MOVPE[21], we added DEZn during growth of the nominal i-segment comprising the QDiscs. Detailed growth parameters can be found in Ref. [19]. The InAsP QDiscs were grown by turning off the trimethylindium (TMIn) flow and replacing the phosphine (PH$_3$) with arsine (AsH$_3$) for 2 s (single QDiscs) / 1 s (20 QDiscs). Scanning transmission electron microscopy (STEM) showed the single discs / multiple discs to be about 8 nm / 5 nm thick, with a composition of about 55% As / 45% P based on point scan measurements.

To make the NW arrays into devices, first, an oxide bi-layer of 50 nm SiO$_x$, then 5 nm Al$_2$O$_3$ was grown by ALD. A photoresist (S1818) planarization layer was spin-coated, before being back-etched via reactive ion etching (RIE) to expose 200-250 nm of the NW tips. The oxide bi-layer at the tip of the NWs, and the gold catalyst particles were subsequently removed by etching. After hard-baking the resist layer, 800×800 μm$^2$ device areas were defined by UV lithography including sputtering of 100 nm ITO as top contact and evaporation of two layers of 20 nm Ti and 400 nm Au for bond pad formation. Details of the processing steps, schematically shown in Figure 2, can be found in Ref. [19].

Initial I-V test measurements were performed at 300 K using a Cascade 11000B probe station in conjunction with a Keithley 4200 semiconductor characterization system. After mounting the samples on DIL-14 holders and bonding, PC measurements were made using a Bruker Vertex 80v Fourier transform infrared (FTIR) spectrometer housing an integrated variable temperature Janis PTSHI-950-FTIR pulse-tube closed-cycle cryostat. The PC response of the devices was characterized using a $CaF_2$ beamsplitter with a quartz lamp for the interband response, or a KBR beam-splitter with MIR source for intersubband measurements. The spectrometer was evacuated to avoid any influence of the absorption lines of air. The modulated (~7.5 kHz) PC was amplified using a Keithley 428 programmable current amplifier. The I-V measurements were recorded with a Keithley 2636 source-meter. The photoluminescence measurements performed using a typical micro-PL setup. A 532 nm green laser from Cabolt was used as excitation source. The sample was mounted on a sample holder integrated within a helium-cooled 5K flow cryostat. Emitted light was collected through an optical microscope, dispersed through a spectrometer and detected by an iDus InGaAs 1.7 μm camera.

**Corresponding Author**

*Email: hakan.pettersson@hh.se

**ACKNOWLEDGEMENTS**

The authors acknowledge financial support from NanoLund, the Swedish Research Council, the Swedish National Board for Industrial and Technological Development, the Knut and Alice Wallenberg Foundation (project 2016.0089), the Swedish Foundation for Strategic Research and the Swedish Energy Agency, the Erik Johan Ljungberg Foundation, and the Carl Trygger Foundation. This project has received funding from the European Union's Horizon 2020 research and innovation program under grant agreement No 641023 (NanoTandem). The authors also acknowledge Mats-Erik Pistol and Federico Capasso for fruitful discussions.

# Supplementary Information

# Intersubband Quantum Disc-in-Nanowire Photodetectors with Normal-incidence Response in the Long-wavelength Infrared

*Mohammad Karimi,[1,2] Magnus Heurlin,[1,3] Steven Limpert,[1] Vishal Jain,[1,2] Xulu Zeng,[1] Irene Geijselaers,[1] Ali Nowzari,[1] Ying Fu,[4] Lars Samuelson,[1] Heiner Linke,[1] Magnus T. Borgström[1] and Håkan Pettersson[1,2*]*

[1]Solid State Physics and NanoLund, Lund University, Box 118, SE-221 00 Lund, Sweden

[2]Department of Mathematics, Physics and Electrical Engineering, Halmstad University, Box 823, SE-301 18 Halmstad, Sweden

[3]Sol Voltaics AB, Lund, Sweden

[4]Department of Applied Physics, Royal Institute of Technology (KTH), Science for Life Laboratory, SE-171 21 Solna, Sweden

## Micro-PL

For micro-PL (µ-PL) studies, single NWs were broken off from the as-grown substrates and mechanically transferred onto Cr-coated silicon substrates. Figure S1 shows the μ-PL of a NW comprising a single QDisc recorded at 4 K for different laser excitation intensity. Taking into account the bandgap bowing parameter of 0.1 eV for WZ InAs at 4 K[24], the estimated bandgap of InAs$_{0.55}$P$_{0.45}$ with 5% variation in As composition would be 0.86-0.96 eV. For the lowest laser intensity of 0.01P, where P is the maximum intensity of 2 W/cm$^2$, a relatively broad peak can be observed at about 0.90 eV. This broad peak contains several peaks in the range from 0.90 to 1.0 eV, reflecting complex recombination processes via multiple states inside the disc. With increasing laser intensity, the μ-PL spectrum is broadened and blue-shifted, reaching about 1.10 eV at the highest laser intensity. Possible explanations for such a spectral behavior could be state-filling, combined with developing type-II transitions from the conduction band of the InP NW to the valence band of the InAsP QDisc. Similar spectral features have recently been reported for InP/InAs/InP core-multishell NWs.[36] Moreover, the non-abrupt interfaces between the QDisc and InP with smooth potential variations would result in broadened PL signals.

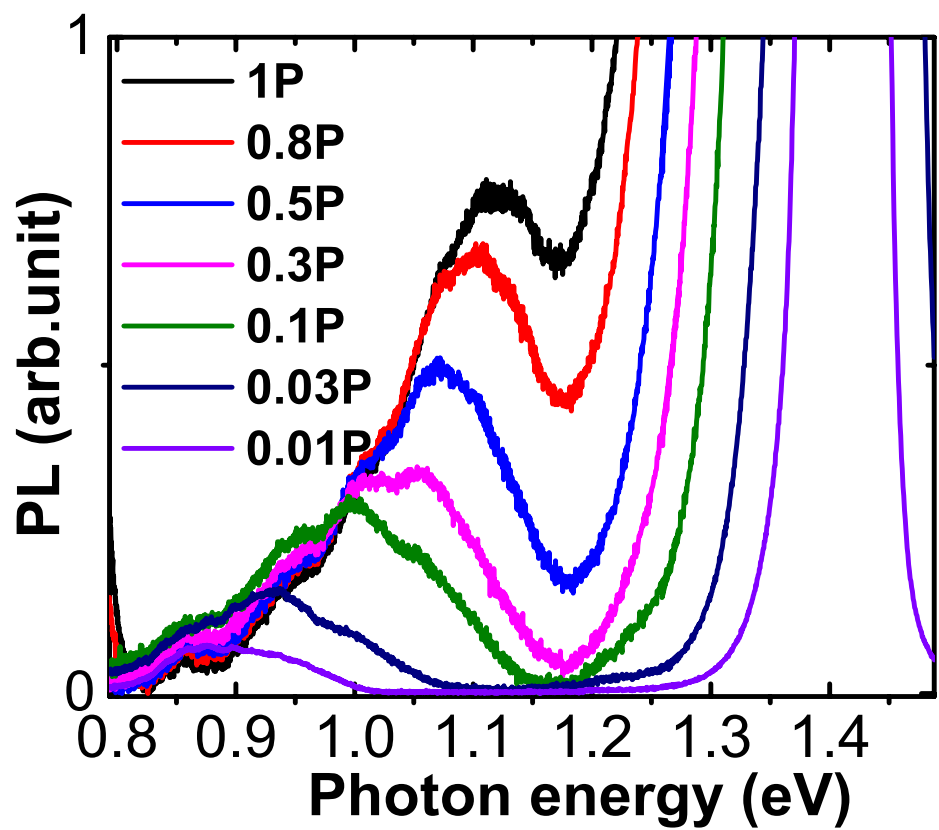

**Figure S1.** μ-PL spectra of a single NW comprising a single QDisc recorded at different laser intensity. The maximum intensity P amounts to 2 W/cm$^2$

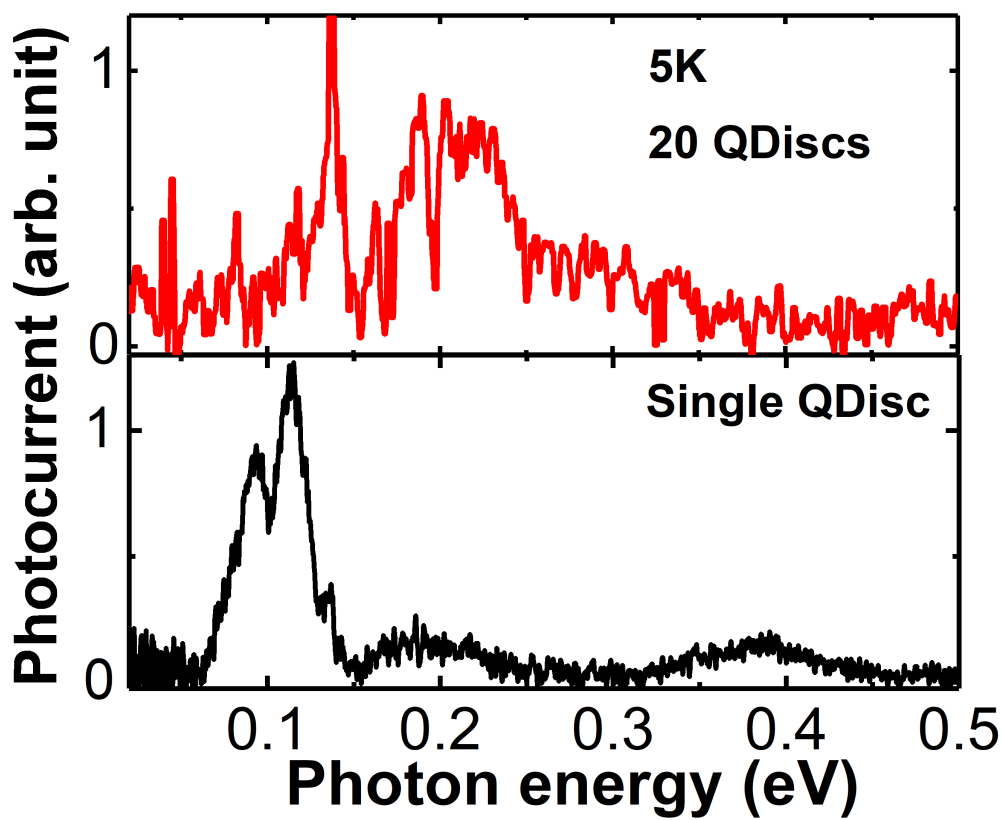

**Figure S2.** The upper panel shows the spectrally-resolved PC for a fully processed 20 QDiscs-in-NW photodetector with growth time of 1 s leading to thin QDiscs, about 4-6 nm in thickness. The lower panel shows the corresponding single QDisc-in-NW detector with thicker 7-8 nm discs (bottom panel). The spectrum in the lower panel is taken from Figure 4a.